\documentclass[journal, 10pt, final, twocolumn, comsoc]{IEEEtran}
\usepackage{cite}
\usepackage{amsmath}
\usepackage{graphicx, subcaption}
\usepackage{subcaption}
\usepackage{float}
\floatstyle{plaintop}
\restylefloat{table}
\usepackage[justification=centering]{caption}
\usepackage{amssymb}
\usepackage{amsmath}
\usepackage{filecontents}
\usepackage{lipsum}
\usepackage{amsthm}
\usepackage{algorithmic,cite}
\usepackage{color}
\usepackage{authblk}
\usepackage[linesnumbered,ruled]{algorithm2e}
\usepackage{enumitem} 

\begin{document}
	\title{Low-Dimensionality of Noise-Free RSS and Its Application in Distributed Massive MIMO}
	\author{K. N. R. Surya Vara Prasad, Ekram Hossain, and Vijay K. Bhargava
	\thanks{K. N. R. S. V. Prasad and V. K. Bhargava are with the Department of Electrical and Computer Engineering at the University of British Columbia, 5500 - 2332 Main Mall, Vancouver, BC V6T 1Z4, Canada (email: surya@ece.ubc.ca; vijayb@ece.ubc.ca). E. Hossain is with the Department of Electrical and Computer Engineering, University of Manitoba, 75A Chancellor's Circle, Winnipeg, MB R3T 5V6, Canada (email: ekram.hossain@umanitoba.ca).}
	}
	\maketitle
	\begin{abstract}
	We examine the dimensionality of noise-free uplink received signal strength (RSS) data in a distributed multiuser massive multiple-input multiple-output system. Specifically, we apply principal component analysis to the noise-free uplink RSS and observe that it has a low-dimensional principal subspace. We make use of this unique property to propose RecGP - a reconstruction-based Gaussian process regression (GP) method which predicts user locations from uplink RSS data. Considering noise-free RSS for training and noisy test RSS for location prediction, RecGP reconstructs the noisy test RSS from a low-dimensional principal subspace of the noise-free training RSS. The reconstructed RSS is input to a trained GP model for location prediction. Noise reduction facilitated by the reconstruction step allows RecGP to achieve lower prediction error than standard GP methods which directly use the test RSS for location prediction.
	\end{abstract}
	
\section{Introduction}	
Massive  multiple-input multiple-output (MIMO) has attracted great attention recently, due to the multifold gains in spectral and energy efficiency it can offer \cite{mm} \cite{ee}. In a distributed massive MIMO (DM-MIMO) system, a large number of base station (BS) antennas are distributed over a service area to cater to multiple users simultaneously on the same-time frequency resource \cite{dm_mimo}. When a user transmits on the uplink, each BS antenna records its own received signal strength (RSS) value and a large vector of RSS values becomes available at the BS. Since the  RSS vectors can be very large, we examine whether they span a low-dimensional principal subspace. To this end, we apply principal component analysis (PCA) \cite{pca_tut} on multiple noise-free RSS vectors and observe that they indeed span a low-dimensional principal subspace. 


As a motivating use-case of the above property, we propose RecGP - a reconstruction method based on Gaussian process regression (GP) \cite{gpr_book} to predict user locations from noisy uplink RSS vectors. We consider a scenario where noise-free RSS is available for training the GP, but only noisy RSS of the test user is available for predicting its location. RecGP reduces the noise present in the test RSS vectors by reconstructing them from a low-dimensional principal subspace of the noise-free training RSS. This noise reduction allows RecGP to achieve lower prediction error than standard GP methods which directly use the test RSS vectors for location prediction.

Few authors have studied system design \cite{dm_mimo} \cite{dm_mimo2} and resource utilization \cite{dm_mimo3} in DM-MIMO, but the low-dimensionality aspect of RSS in DM-MIMO has not been explored. We apply PCA to noise-free uplink RSS in DM-MIMO and report for the first time that it has a low-dimensional principal subspace. Recently, the authors in \cite{rss_gp5} have proposed a standard GP method for location prediction in DM-MIMO, but with noisy RSS for both training and prediction. In contrast, we consider noise-free RSS for training and propose a new GP method which exploits the low-dimensionality of noise-free training RSS to achieve lower prediction error than standard GP.

\subsubsection*{Notation} We use boldface small and capital letters for vectors and matrices respectively. The notations $[\mathbf{a}]_i$, $[\mathbf{A}]_i$, and $[\mathbf{A}]_{ij}$ refer to the element $i$ in vector $\mathbf{a}$, column $i$ in matrix $\mathbf{A}$, and the element $(i,j)$ in matrix $\mathbf{A}$, respectively. Overhead symbols $\widetilde{(.)}$ and $\widehat{(.)}$ refer to training and test data, respectively, with an additional superscript $(.)^*$ if the data is noise-free. 

\section{System Description} \label{sec_sys_model}
We consider multiuser transmissions in a DM-MIMO system, where $K$ users transmit radio signals to $M$ distributed BS antennas simultaneously and on the same time-frequency resource. The BS antennas are connected to a computing unit (CU) via high-speed backhaul to offload all the computational load. When the $K$ users transmit on the uplink, each BS antenna records its own RSS value and an $M \times 1$ vector of RSS values becomes available at the CU for further processing.

\begin{figure*}
	\begin{subfigure}{.24\textwidth}
		\centering
		\includegraphics[width=1\linewidth]{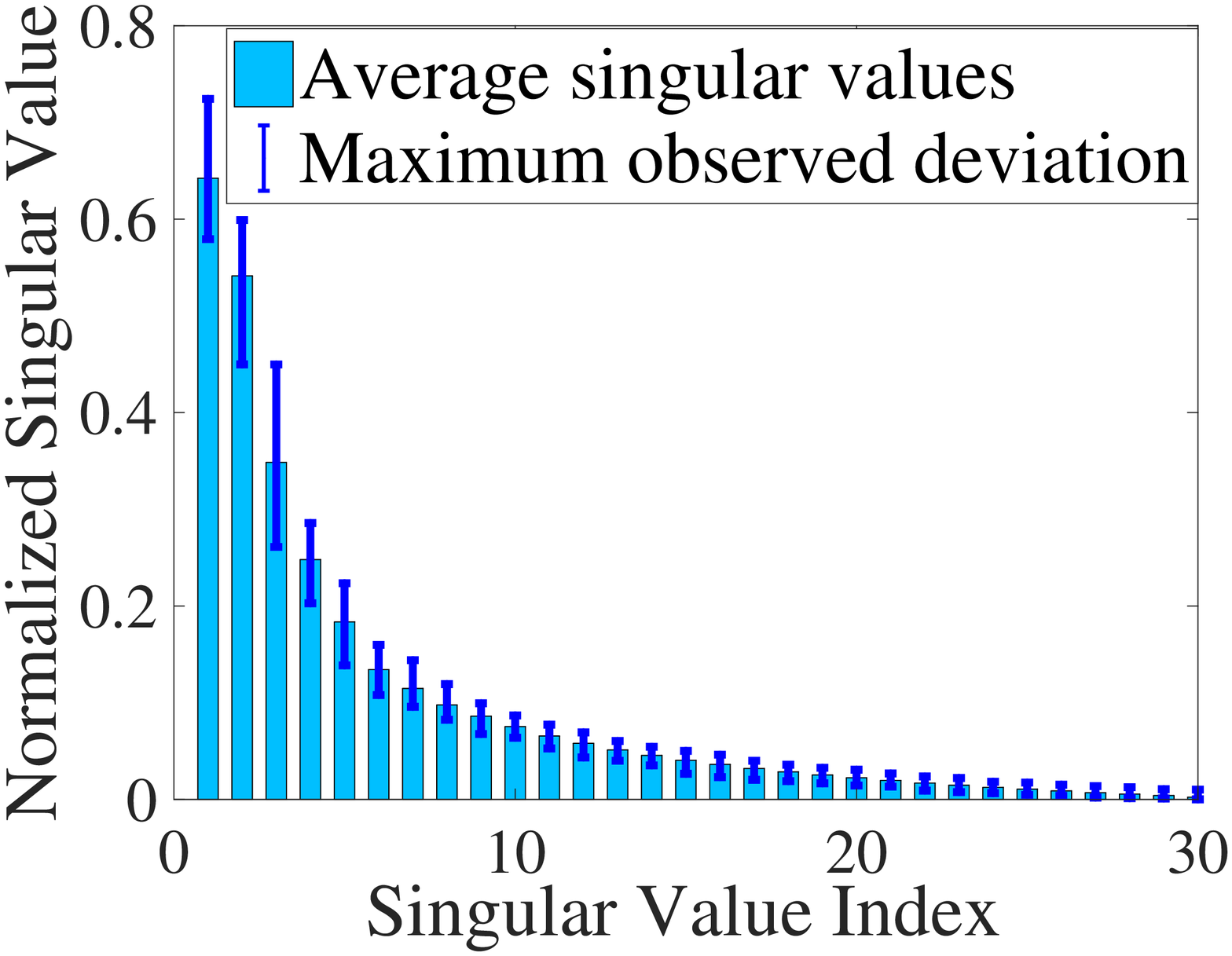}
		\captionsetup{justification=centering}
		\caption{$M=30$}
		\label{figure_1a}
	\end{subfigure}%
	\begin{subfigure}{.24\textwidth}
		\centering
		\includegraphics[width=1\linewidth]{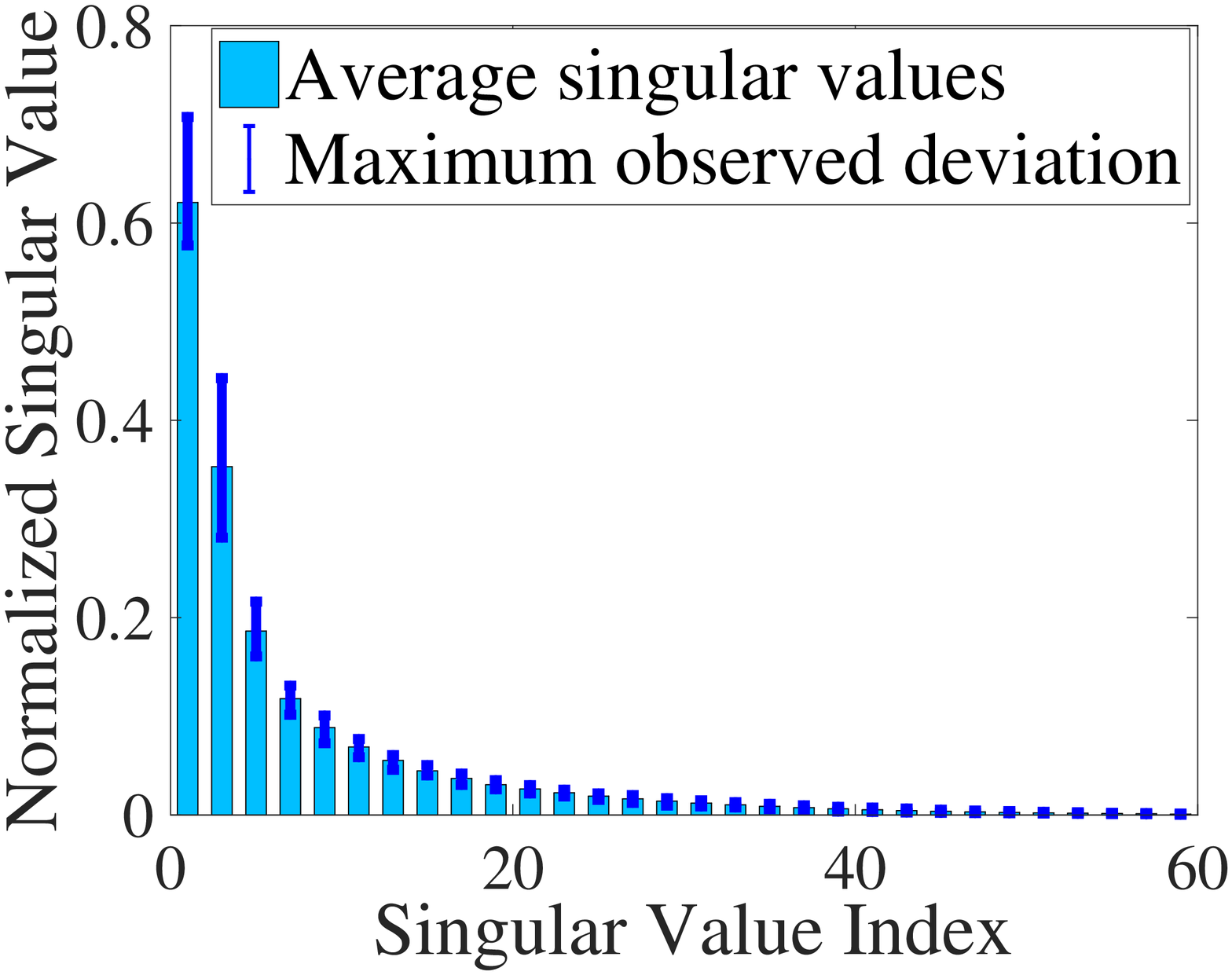}
		\captionsetup{justification=centering}
		\caption{$M=60$}
		\label{figure_1b}
	\end{subfigure}%
	\begin{subfigure}{.24\textwidth}
		\centering
		\includegraphics[width=1\linewidth]{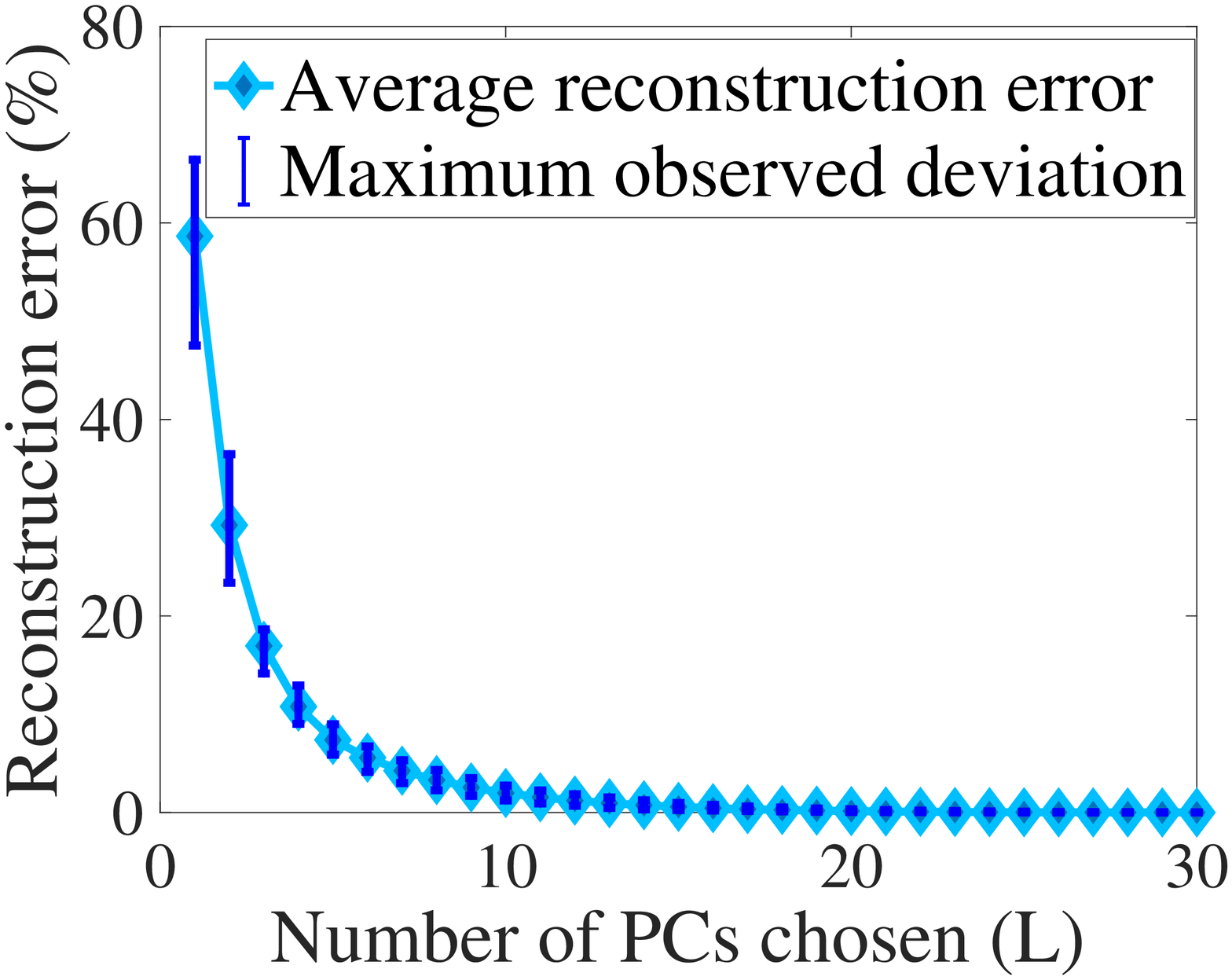}
		\captionsetup{justification=centering}
		\caption{$M = 30$} 
		\label{figure_1c}
	\end{subfigure}
	\begin{subfigure}{.24\textwidth}
		\centering
		\includegraphics[width=1\linewidth]{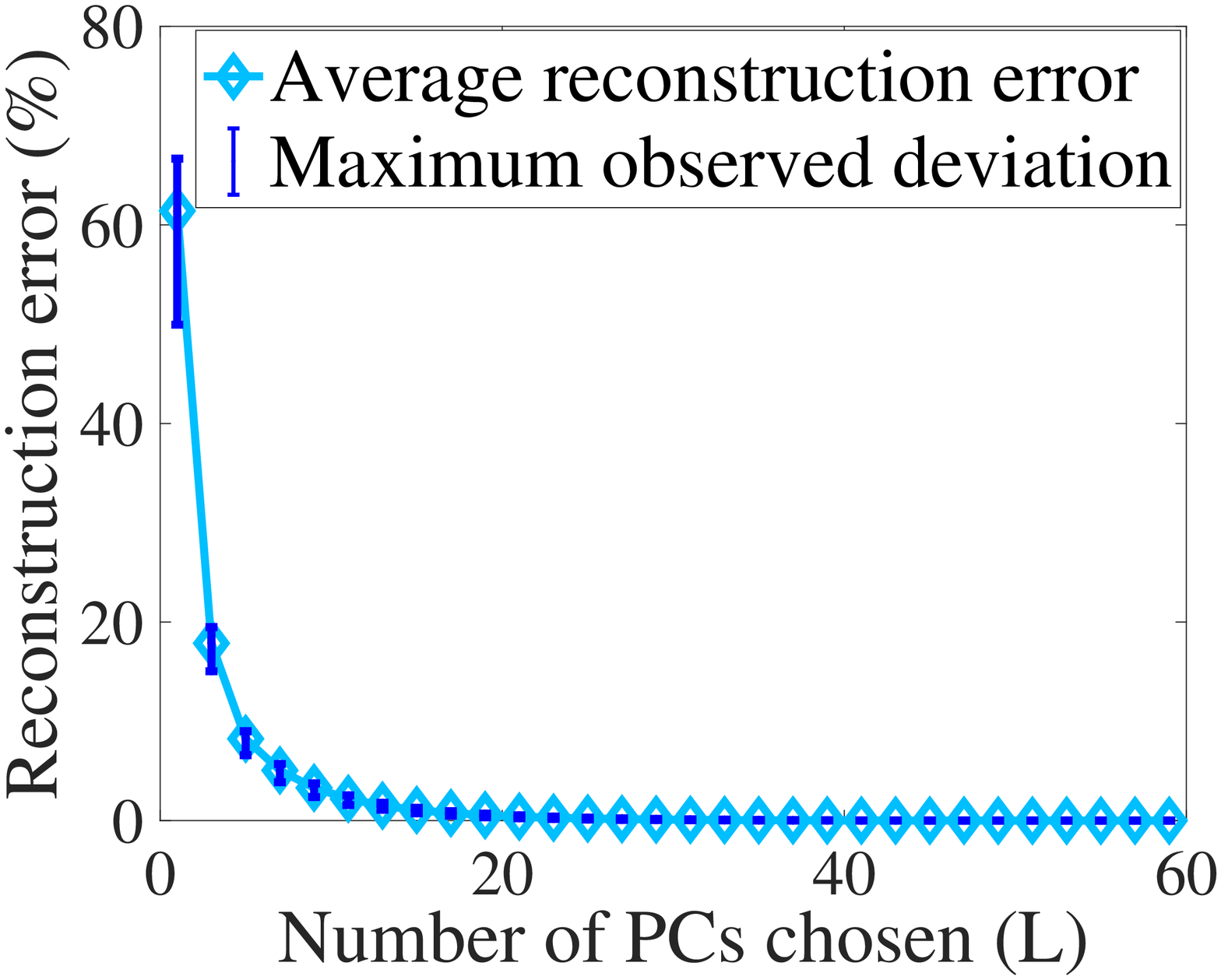}
		\captionsetup{justification=centering}
		\caption{$M=60$}
		\label{figure_1d}
	\end{subfigure}%
	\caption{Plots of the singular values and reconstruction errors (with the first $L$ PCs) for $M = 30$ and $M = 60$, averaged over $200$ different noise-free RSS matrices. In Figs. \ref{figure_1a}-\ref{figure_1b}, we observe that the first few singular values explain most of the energy contained in $\mathbf{P}^*$. In 
	Figs. \ref{figure_1c}-\ref{figure_1d}, we observe that the first few PCs  reconstruct more than $95 \%$ of the data contained in $\mathbf{P}^*$.}
	\label{figure_1}
\end{figure*}

\subsection{Uplink Transmissions in DM-MIMO} Let $\mathbf{s}_k$ be the symbol vector transmitted by user $k$ and $\rho$ be the transmission power of each scheduled user. The symbol vector $\mathbf{r}_m$ received by the BS antenna $m$ is given by
\begin{equation} \label{eqRxsignal}
\mathbf{r}_m = \sqrt{\rho} \sum_{k=1}^{K} h_{km} \mathbf{s}_k + \mathbf{z}_m^{\text{rx}},
\end{equation}
where $h_{mk} = q_{mk} \sqrt{\beta_{mk}} $ is the flat-fading uplink channel gain with $q_{mk}$ and $\beta_{mk}$ being small-scale and large-scale fading coefficients, and $\mathbf{z}_m^{\text{rx}} \sim \mathcal{N} (\mathbf{0}, \sigma_{\text{rx}}^2 \mathbf{I})$ is the additive white Gaussian noise vector. We assume that the coefficients $q_{mk}$ are independent and identically distributed complex normal random variables, i.e., $q_{mk} \sim \mathcal{CN} (0,1)$, and model $\beta_{mk}$ as
\begin{equation} \label{eqChannel}
\begin{aligned}
\beta_{mk} = l_0 d_{mk}^{-\eta} 10^{\frac{z_{mk}^{\text{sh}}}{10}}
\end{aligned}
\end{equation}
where $d_{mk}$ is the distance between the user $k$ and BS antenna $m$, $l_0$ is the reference path-loss at a distance $d_0$, $\eta$ is the path-loss exponent, and $z_{mk}^{\text{sh}} \sim \mathcal{N} (0,\sigma_{sh}^2 )$ is channel power gain due to shadowing. 

\subsection{Obtaining Noise-free RSS} 
 The RSS $p_{mk}$ of each user $k$ should be extracted from the sum-RSS $||\mathbf{r}_m||^2$. This can be done if the uplink vectors $\{\mathbf{s}_k\}$ in (\ref{eqRxsignal}) are mutually orthogonal and are known at the BS. For example, $\{\mathbf{s}_k\}$ can be pilot sequences used in signal detection \cite{cf_mimo}. The RSS $p_{mk}$ of user $k$ can then be obtained from (\ref{eqRxsignal}) as
\begin{equation} \label{eq_pmk}
p_{mk} = \rho \beta_{mk} |q_{mk}|^2.
\end{equation}
Observe from (\ref{eq_pmk}) that the extracted RSS values can be noisy due to small-scale fading and shadowing effects. While the small-scale fading can be averaged out over multiple timeslots, shadowing can be spatially averaged out if we have prior access to the user's location. For example, the BS can record the RSS averaged over nearby locations with approximately the same user-to-BS distance. When both multi-timeslot and spatial averaging are employed, we can obtain the noise-free RSS $p_{mk, \text{dB}}^*$ of each user $k$ (in dB scale) from (\ref{eqChannel}) and (\ref{eq_pmk}), as
\begin{equation} \label{eq_rss_dB}
\begin{aligned}
p_{mk, \text{dB}}^* & = p_{0, \text{dB}} - 10 \eta \log_{10} (d_{mk}), \\
\end{aligned}
\end{equation}
\noindent where $p_{0, \text{dB}} = 10 \log_{10} (\rho l_0) $ is the reference RSS at distance $d_0$ and a superscript $(.)^*$ is given to $p_{mk, \text{dB}}$ to highlight that it is noise-free. The BS can then form an $M \times 1$ noise-free RSS vector $\mathbf{p}_k^*$ for each user $k$ such that 
\begin{equation}
\mathbf{p}_k^* = [p_{1k, \text{dB}}^* \, \, p_{2k, \text{dB}}^* \, \dots \, p_{Mk, \text{dB}}^*]^T.
\end{equation}
Following the same procedure, the BS can extract noise-free RSS vectors for $N$ different user locations and accumulate them into an $N \times M$ noise-free RSS matrix $\mathbf{P}^*$, such that
\begin{equation} \label{nf_rss}
\mathbf{P}^* = [\mathbf{p}_1^*\, \mathbf{p}_2^*\, \dots \mathbf{p}_{N}^*]^T.
\end{equation}

\section{Low-Dimensionality of the Noise-Free RSS}
We take a simulation approach to demonstrate that the noise-free RSS matrix $\mathbf{P}^*$ in (\ref{nf_rss}) has a low-dimensional principal subspace. We consider two example scenarios with $ M=30$ and $M = 60$ BS antennas distributed randomly over a $500$m $\times 500$m service area. A sample noise-free RSS matrix $\mathbf{P}^*$ is built by choosing $N = 1000$ locations distributed randomly in the service area and using (\ref{eq_rss_dB}) with parameters as per Table \ref{table_pathloss} to generate the noise-free RSS vectors. This is repeated to build $200$ different RSS matrices each for $M = 30$ and $M = 60$.

We now decompose each sample matrix $\mathbf{P}^*$ into three parts via singular value decomposition \cite{pca_tut} to obtain 
\begin{equation} \label{eq_svd}
\mathbf{P}^* = \mathbf{U} \mathbf{D} \mathbf{V}^T, 
\end{equation}
where columns of the orthogonal matrices $\mathbf{U}$ and $\mathbf{V}$ are the left singular and right singular vectors of $\mathbf{P}^*$, and the diagonal elements in $\mathbf{D}$ are the singular values of $\mathbf{P}^*$ arranged in decreasing order. In Fig. \ref{figure_1a} and \ref{figure_1b}, we plot the singular values, averaged over the $200$ different $\mathbf{P}^*$ matrices, for $M = 30$ and $M = 60$, respectively. Error bars represent the maximum observed deviation from average values. For both $M = 30$ and $M = 60$, we notice that the first few singular values represent most of the energy contained in the $\mathbf{P}^*$. 

For further insight, we study the error incurred upon reconstructing the noise-free RSS matrices from the subspace spanned by the first $L$ principal components (PCs).  Using truncated SVD \cite{pca_tut}, we can reconstruct each sample matrix $\mathbf{P}^*$ from its first $L$ PCs as $\mathbf{U}^{[L]} \mathbf{D}^{[L]} \mathbf{V}^{[L]T}$, where $\mathbf{U}^{[L]} $and $\mathbf{V}^{[L]}$ are matrices formed by the first  $L$ columns of $\mathbf{U}$ and $\mathbf{V}$, respectively, and $\mathbf{D}^{[L]}$ is the diagonal matrix formed by the first $L$ singular values of $\mathbf{P}^*$. The reconstruction error $||\mathbf{P}^* - \mathbf{U}^{[L]} \mathbf{D}^{[L]} \mathbf{V}^{[L]T}||^2$, averaged over the $200$ different $\mathbf{P}^*$ matrices, is plotted in Figs. \ref{figure_1c} and \ref{figure_1d} against the number of chosen PCs $L$. We observe that for both $M = 30$ and $M=60$, the first few PCs are consistently able to reconstruct more than $95\%$ of the data contained in $\mathbf{P}^*$. Similar plots are observed for $M$ ranging from $30$ to $100$. These plots show that we can form a low-dimensional principal subspace of the noise-free RSS by combining the first $L$ PCs of $\mathbf{P}^*$, with $L$ chosen to keep the reconstruction error below a certain threshold (for example, $5\%$). In the next section, we present a motivating use-case which exploits the low-dimensionality of this principal subspace to predict user locations in DM-MIMO.

\section{RecGP: A GP Method for Location Prediction}
We propose RecGP, which is a reconstruction-based GP method to predict user locations from uplink RSS vectors in DM-MIMO. As in standard GP \cite{gpr_book}, we train a GP model with RSS vectors for several known user locations. The trained GP model, when input with the RSS vector of a test user, outputs an estimate of the test user's location.
We consider noise-free RSS for training the GP because small-scale fading can be averaged out over multiple time slots and shadowing can be spatially averaged out using our access to the training user locations. In contrast, we consider the test RSS vectors as noisy due to shadowing. This is because, although time-averaging can mitigate small-scale fading, we do not have access to the test user's location and are therefore unable to spatially average out the shadowing noise present in test RSS.

While the standard GP directly inputs the test RSS vectors to a trained GP model for location prediction, RecGP first reduces the noise in test RSS vectors by reconstructing them from a low-dimensional principal subspace of the noise-free training RSS. The reconstructed RSS vectors are input to a trained GP model for location prediction. Details of the training and prediction phases in RecGP are presented next, with focus on $x-$coordinates\footnotemark of the users.
\footnotetext{The presented method is equally valid for $y-$coordinates as well.} 

\subsubsection{Training Phase}
We train a GP model to learn the function $f_x(.)$ which maps the RSS vector $\mathbf{p}_k$ of any user $k$ to its $x-$coordinate $x_k$ such that $x_k = f_x(\mathbf{p}_k), \, \forall x_k$. At the core of GP methods is the assumption that any finite set of realizations of the function to be learned, i.e., $f_x(.)$, follow a zero-mean Gaussian distribution with a covariance matrix $\mathbf{\Phi}$ whose elements are given by a user-defined function $\phi(.)$ \cite{gpr_book}. In short, we say $f_x(.) \sim \mathcal{GP} (0, \phi(.))$. The function $f_x(.)$ is fully specified by $\phi(.)$ because a Gaussian distribution is fully specified by its mean and variance. Functionally, $\phi(.)$ models the covariance of $x-$coordinates of any two users $i$ and $j$ in terms of their RSS vectors $\mathbf{p}_i$ and $\mathbf{p}_j$. We choose $\phi(.)$ as \cite{cov_gp}
\begin{equation}\label{eq_cov_model}
\begin{aligned}
\phi(\mathbf{p}_i, \mathbf{p}_j) = \alpha \exp((\mathbf{p}_i - \mathbf{p}_j)^T \mathbf{B}^{-1} (\mathbf{p}_j - \mathbf{p}_i)) + \gamma \mathbf{p}_i^T \mathbf{p}_j,  \\
\end{aligned}
\end{equation}
where the exponential term and the inner product terms model the dependence of $\phi(\mathbf{p}_i, \mathbf{p}_j)$ on the distance ($\mathbf{p}_i - \mathbf{p}_j$) and the actual RSS $\mathbf{p}_i$ and $\mathbf{p}_j$, respectively. The model in (\ref{eq_cov_model}) introduces a free-parameter vector $\boldsymbol{\theta} = [\alpha; \, [\mathbf{B}]_{11} ; \, [\mathbf{B}]_{22} ;\, \dots \, [\mathbf{B}]_{MM}; \, \gamma]$. We learn $\boldsymbol{\theta}$ via maximum-likelihood of the vector $\widetilde{\mathbf{x}} = [\widetilde{x}_1 \, \widetilde{x}_2 \, \dots \, \widetilde{x}_{\widetilde{N}}]^T$ of $\widetilde{N}$ training user  $x-$coordinates, as
\begin{equation}\label{eq_theta_opt}
\begin{aligned}
\bar{\boldsymbol{\theta}} & = \underset{\boldsymbol{\theta}}{\arg \max} \log (p(\widetilde{\mathbf{x}} | \widetilde{\mathbf{P}}^*, \boldsymbol{\theta})).
\end{aligned}
\end{equation}
In (\ref{eq_theta_opt}), $\bar{\boldsymbol{\theta}}$ is the learned vector $\boldsymbol{\theta}$, $\widetilde{\mathbf{P}}^* = [\widetilde{\mathbf{p}}_1^*\, \widetilde{\mathbf{p}}_2^*\, \dots \widetilde{\mathbf{p}}_{\widetilde{N}}^*]^T$ is the noise-free training RSS matrix, and the distribution of $(\widetilde{\mathbf{x}} | \widetilde{\mathbf{P}}^*, \boldsymbol{\theta})$ follows from the GP assumption $f_x(.) \sim \mathcal{GP} (0, \phi(.))$ as \cite{gpr_book} 
\begin{equation} \label{eq_trainingdist}
\begin{aligned}
\widetilde{\mathbf{x}} | \widetilde{\mathbf{P}}^*, \boldsymbol {\theta} & \sim \mathcal{N} ( \mathbf{0}, \widetilde{\mathbf{\Phi}}^*), \text{ where} \\
[\widetilde{\mathbf{\Phi}}^*]_{n,n'} &= \phi(\widetilde{\mathbf{p}}_n^*, \widetilde{\mathbf{p}}^*_{n'}), \quad n, n' = 1, \dots, \widetilde{N}. 
\end{aligned}
\end{equation}
The problem in (\ref{eq_theta_opt}) is non-convex, but can be solved for local optimum using gradient ascent methods such as conjugate gradient \cite{gpr_book}. Learning $\boldsymbol{\theta}$ completes the training phase because the $x-$coordinate function $f_x(.)$ is fully specified by $\phi(.)$.

\subsubsection{Reconstruction Phase}
Let $\widehat{\mathbf{x}} = [\widehat{x}_1 \, \widehat{x}_2 \, \dots \, \widehat{x}_{\widehat{N}}]^T$ be the $\widehat{N} \times 1$ vector of the test users' $x-$coordinates that we should predict, and $\widehat{\mathbf{P}} = [\widehat{\mathbf{p}}_1\, \widehat{\mathbf{p}}_2\, \dots \widehat{\mathbf{p}}_{\widehat{N}}]^T$ be the corresponding matrix of noisy test RSS vectors. If we combine the first $L$ PCs of the noise-free training RSS $\widetilde{\mathbf{P}}^*$ to form a low-dimensional principal subspace, we can reconstruct the noisy test RSS $\widehat{\mathbf{P}}$ from this principal subspace as follows \cite{pca_tut}:

\begin{equation} \label{eq_noisyrss_recon}
\widehat{\mathbf{P}}^{(rec)} = \widehat{\mathbf{P}} \mathbf{V}^{[L]} \mathbf{V}^{[L]T}, 
\end{equation}
where $\mathbf{V}^{[L]}$ is the matrix formed by the first $L$ right singular vectors of $\widetilde{\mathbf{P}}^*$. Eq. (\ref{eq_noisyrss_recon}) inherently facilitates noise reduction because the noisy test RSS vectors are projected onto a subspace spanned by noise-free RSS. The reconstructed RSS matrix $\widehat{\mathbf{P}}^{(rec)}$ is input to the trained GP for predicting $\widehat{\mathbf{x}}$.

\subsubsection{Prediction Phase}
As per the GP assumption $f_x(.) \sim \mathcal{GP} (0, \phi(.))$, the training and test vectors $\widetilde{\mathbf{x}}$ and $\widehat{\mathbf{x}}$ are jointly Gaussian distributed. Conditioning on this joint distribution gives the predictive distribution of the $x-$coordinate $[\widehat{\mathbf{x}}]_n$ of a test user $n$ whose reconstructed RSS vector is $\widehat{\mathbf{p}}_n^{\text{(rec)}}$, as \cite{gpr_book}

\begin{equation} \label{eq_pred_recgp_user}
\begin{aligned}
& \text{$[\widehat{\mathbf{x}}]_n$} | \widetilde{\mathbf{x}}, \widetilde{\mathbf{P}}, \widehat{\mathbf{p}}_n^{\text{(rec)}}   \sim \mathcal{N} ([\widehat{\boldsymbol{\mu}}_{x}]_n, [\widehat{\textbf{C}}_{x}]_{nn}), \text{ where } \\
& [\widehat{\boldsymbol{\mu}}_{x}]_n  = \sum_{i=1}^{\widetilde{N}} \phi(\widehat{\mathbf{p}}_n^{\text{(rec)}}, \widetilde{\mathbf{p}}_i^*) [(\widetilde{\mathbf{\Phi}}^*)^{-1} \widetilde{\mathbf{x}}]_i, \text{ and} \\
& [\widehat{\textbf{C}}_{x}]_{nn} = \phi(\widehat{\mathbf{p}}_n^{\text{(rec)}}, \widehat{\mathbf{p}}_n^{\text{(rec)}}) - \sum_{i=1}^{\widetilde{N}} \sum_{j=1}^{\widetilde{N}} \{ \phi(\widehat{\mathbf{p}}_n^{\text{(rec)}}, \widetilde{\mathbf{p}}_i^*) \\
& \quad \quad \quad \quad  [(\widetilde{\mathbf{\Phi}}^* )^{-1}]_{ij} \phi(\widetilde{\mathbf{p}}_j^*, \widehat{\mathbf{p}}_n^{\text{(rec)}}) \}.
\end{aligned}
\end{equation}
 In (\ref{eq_pred_recgp_user}), $[\widehat{\boldsymbol{\mu}}_{x}]_n$ and $[\widehat{\textbf{C}}_{x}]_{nn}$ are the predicted mean and variance of the $x-$coordinate $[\widehat{\mathbf{x}}]_n$ of the test user $n$. Since the mean of a Gaussian distribution is also its mode, $[\widehat{\boldsymbol{\mu}}_{x}]_n$ gives us the maximum-a-posteriori (MAP) estimate of $[\widehat{\mathbf{x}}]_n$. Also, $[\widehat{\textbf{C}}_{x}]_{nn}$ gives us the confidence interval $[\widehat{\boldsymbol{\mu}}_{x}]_n \pm 2 \sqrt {[\widehat{\textbf{C}}_{x}]_{nn}}$  on choosing $[\widehat{\boldsymbol{\mu}}_{x}]_n$ as the predicted estimate of $[\widehat{\mathbf{x}}]_n$. RecGP achieves lower prediction error than standard GP, thanks to the noise reduction from reconstruction of the test RSS vectors.


\begin {table}
\caption{Simulation values as per the urban micro model in 3GPP TR 25.814 \cite{3gpp} \label{table_pathloss} }
\fontsize{12}{12}
\centering 
\begin{tabular}{| l | l | l |}
	\hline
	{Simulation parameter} & {Value} \\ \hline \hline
	User transmit power ($\rho$) & $21$dBm ($125$mW) \\ \hline
	Reference distance ($d_0$) & $10$m  \\ \hline 
	Reference pathloss ($l_0$) & $-47.5$dB \\ \hline
	Path-loss exponent ($\eta$) & $ 0, \text{ if } d_{mk} < 10\text{m}$ \\ 
	& $ 2, \text{ if } 10\text{m} \leq d_{mk} \leq 45\text{m}$ \\
	& $6.7, \text{ otherwise}$ \\ \hline
\end{tabular}
\end {table}

\section{Simulation Studies} \label{sec_sim}
We consider an example DM-MIMO setup with $M = 30$ BS antennas, $\widetilde{N} = 400$ training user locations, and $\widehat{N} = 25$ test user locations, all distributed uniformly over a service area of $500$m $\times$ $500$m. For training, we generate a noise-free RSS matrix $\widetilde{\mathbf{P}}^*$ using (\ref{eq_rss_dB}) with parameters given by Table \ref{table_pathloss}. We then solve the log-likelihood maximization problem in (\ref{eq_theta_opt}) using conjugate gradient method \cite{gpr_book}. Multiple trials are run with random initial values to avoid choosing a bad local optimum. For the prediction phase, we generate test RSS matrices $\widehat{\mathbf{P}}$ using (\ref{eq_rss_dB}) with an additional shadowing term $z_{mk}^{\text{sh}} \sim \mathcal{N} (0, \sigma_{\text{sh}}^2)$ and other parameters as per Table \ref{table_pathloss}. We measure prediction performance in terms of the root mean squared error (RMSE) between the actual coordinates ($[\widehat{\mathbf{x}}]_n$, $[\widehat{\mathbf{y}}]_n$) of the test users and their estimates  ($[\widehat{\boldsymbol{\mu}}_{x}]_n$, $[\widehat{\boldsymbol{\mu}}_{y}]_n$). The standard GP (SGP) method, which predicts user locations using test RSS vectors without reconstruction, serves as the baseline for comparison. 

\begin{figure}
	\begin{center}
		\includegraphics[scale=0.45] {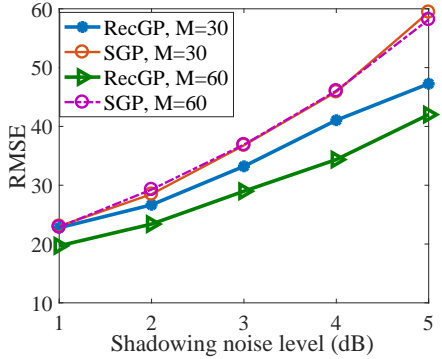}
		\caption{Average RMSE performance for $M = 30$ and $M= 60$. RecGP consistently outperforms SGP, thanks to noise reduction from reconstruction of the test RSS.}
		\label{figure_rmse}
	\end{center}
\end{figure}

In Fig. \ref{figure_rmse}, we plot the RMSE performance of RecGP and SGP, averaged over $200$ Monte-Carlo realizations of the test RSS matrices and the $\widehat{N}$ test user locations, for shadowing noise $\sigma_{\text{sh}}^2$ ranging from $1$dB to $5$dB. To reconstruct the test RSS, we chose $L$ as the number of PCs which most-frequently gave the lowest RMSE among the Monte-Carlo datasets.  For both $M = 30$ and $M = 60$, we observe that RecGP consistently outperforms SGP, thanks to the noise-reduction from projecting the test RSS vectors onto a low-dimensional principal subspace of the noise-free RSS. Also, when the number of antennas is doubled from $M=30$ to $M=60$, we observe that the RMSE performance of RecGP has improved, but there was a negligible impact on SGP. Lastly, we observe that the RMSE of both SGP and RecGP increases with the noise level. Because both the methods are trained with noise-free RSS, they tend to project the noise present in input RSS onto the output location coordinate space.

\begin{figure}
	\begin{center}
		\includegraphics[scale=0.45] {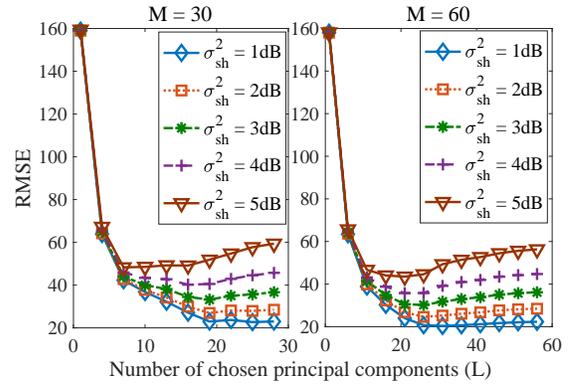}
		\caption{Average RMSE of RecGP vs. number of PCs $L$. RMSE decreases initially, followed by a gradual increase because $L$ introduces a trade-off between the amount of information lost and the amount of noise reduced through reconstruction.}
		\label{figure_rmsevssigmodes}
	\end{center}
\end{figure}

In Fig. \ref{figure_rmsevssigmodes}, we plot the average RMSE performance of RecGP for $M = 30$ and $M = 60$, when the number of chosen PCs $L$ is increased from $1$ to $30$ and $60$, respectively. For very low $L$, the RMSE is very high because we lose most of the information contained in the test RSS through reconstruction. Upon increasing $L$, RMSE decreases initially, attains a minimum level, followed by a gradual increase, with the increase being more prominent for higher noise levels. This is expected, because $L$ introduces a trade-off between the amount of information lost and the amount of noise reduced through the reconstruction procedure. Also, note that the RMSE-minimizing $L$ is different for different noise levels. We therefore choose $L$, for a given noise level, as the number of PCs which most-frequently gives the lowest RMSE among the Monte-Carlo datasets. 

\section{Conclusion}\label{sec_conclusion}
We have applied principal component analysis to the noise-free uplink RSS data in a distributed massive MIMO (DM-MIMO) system and observed that it spans a low-dimensional principal subspace. This interesting property can be exploited for performance improvement in relevant machine learning applications. As a motivating use-case, we have proposed RecGP - a reconstruction-based Gaussian process regression (GP) method which predicts user locations in DM-MIMO from uplink RSS data. When noise-free RSS is available for training, but only noisy RSS of the test user is available for location prediction, RecGP reconstructs the noisy test RSS from a low-dimensional principal subspace of the noise-free training RSS. The reconstructed test RSS is used for location prediction, as opposed to the standard GP method of directly using the test RSS for the same. Simulation studies have confirmed that the reconstruction step has reduced noise in the test RSS and has empowered RecGP to achieve better prediction performance than standard GP.

\end{document}